\documentclass[aps,prl,twocolumn,showpacs,superscriptaddress,showkeys]{revtex4-1}

\usepackage{amssymb}
\usepackage{amsmath}
\usepackage{amsfonts}
\usepackage{graphicx}
\usepackage{color}
\usepackage{xspace}
\usepackage{ulem}
\usepackage{mathtools}
\usepackage{hhline}

\newcommand{\AddrVandy}{Department of Physics and Astronomy, Vanderbilt University, Nashville, TN 37235, USA}

\begin{document}

\title{Photon directional profile from stimulated decay of axion clouds with arbitrary axion spatial distributions}

\author{Liang Chen} \email{bqipd@pm.me}\affiliation{\AddrVandy}
\author{Thomas W. Kephart}    \email{tom.kephart@gmail.com}\affiliation{\AddrVandy}

\date{\today}

\begin{abstract}
We model clusters of axions with spherically symmetric momentum but arbitrary  spatial distributions and study the directional profile of photos produced in their evolution through spontaneous and stimulated decay of axions via the process $a \rightarrow \gamma + \gamma$. Several specific examples are presented. 

\end{abstract}


\maketitle
\section{Introduction}
Axions are copiously produced at the QCD phase transition. A possible way to detect these cosmological axions is through the observation of lasing axion clouds (clumps). If axions are a component of the cold dark matter (CDM), they can form density perturbations in the early Universe. If the over dense regions have high enough number density, then ambient photons from the cosmic microwave background (CMB) or from spontaneous axion decays, can induce stimulated  axion decay within the clumps, i.e., the axions can lase \cite{Kephart:1986vc,Tkachev:1987cd,Kephart:1994uy}.

Besides the initial clumps, other axion structures can form. The initial density perturbations can infall and evolve to form caustics \cite{Sikivie:1997ng} which have complicated geometries. Yet another possibility is that axions can be produced after the formation of primordial black holes (PBHs). Such black holes can be the results of various early universe processes, from cosmic string or domain wall singularities to density perturbations. However they for, if they have sufficient angular, either initially or from mergers, then superradience can occur causing axions to populate an $n,l,m=2,1,1$ hydrogen-like orbit around them if the axion Compton wavelength is comperiable to the PBHs' radius. If the axion density is high enough they can lase \cite{Rosa:2017ury}. The process can saturate, stop and then repeat in what is similar to what has been seen for fast radio bursts (FRBs).

Lasing in the PBH superradience case has so far only been approximated using the sphyerically symmetric model \cite{Rosa:2017ury}. In this work and in \cite{Chen:2020ufn} we point the way to an improving this approximation using multipole expansions of the spatial and momentum space distributions to more closely represent the physical axion distributions expected around a PBH.

\section{Photon angular distribution} 
In \cite{Kephart:1986vc,Kephart:1994uy} nonrelativistic axions of mass $m_a$ were contained in a ball of radius $R$, with a maximum momentum value of $p_{\textrm{\tiny max}} \approx m_a\beta$. Here we allow a non spherically symmetric spatial distribution $X(\theta,\phi)$ to modify the axion clouds model previously studied, with the aim of finding the angular distribution $Y(\theta,\phi)$ of photons resulted from decays of axions, providing that there is some outside constraint (e.g., a gravitational field or self interactions) that can keep the axions  in the initial spatial distribution. For such an axion distribution, assuming it factorizes, the occupation number $f_a(p, r, \theta, t)$ and number densities $n_a(r, \theta, t)$ can be written
\begin{flalign}\label{axionON}
& f_a(p, r, \theta, t) =f_{ac}(t)\Theta(p_{\textrm{\tiny max}}-p)\, \Theta(R-r)  X(\theta)  \\ \nonumber
\end{flalign}
and
\begin{flalign}
&  n_a(r, \theta, t) =\int\frac{d^3p}{(2\pi)^3} f_a(p, r, \theta, t)  \\ \nonumber
&\qquad\qquad = \frac{m_a^3\beta^3}{6\pi^2} f_{ac}(t) \Theta(R-r) X(\theta)  \\ \nonumber 
&\qquad\qquad =n_{ac}(t)\Theta(R-r) X(\theta)   \label{axion-fn}  
\end{flalign}
where we can translate between the two with
\begin{flalign}
 f_{ac}(t) =\frac{6\pi^2}{ m_a^3 \beta^3} n_{ac}(t) .
\end{flalign}
 
Here and elsewhere we use the short hand notation $X(\theta)$ for $X(\theta,\phi)$, likewise for $Y$, $f$ and $n$.

The photons are contained in a ball of radius $R$, a momentum spherical shell of inner and outer radius $k_-  =\frac{m_a\gamma}{2}(1-\beta)$ and $k_+ = \frac{m_a\gamma}{2}(1+\beta) $ respectively \cite{Kephart:1994uy}, where we use $\beta =v/c$.
\begin{flalign}\label{photonON}
& f_\lambda(k, r, \theta, t) = f_{\lambda c}(t) \Theta(k_+-k) \Theta(k-k_-) \Theta(R-r)Y(\theta)    \\ \nonumber
\end{flalign}
and
\begin{flalign}
 &n_\lambda( r, \theta, t)   = \int\frac{d^3k}{(2\pi)^3} f_\lambda(k, r, \theta, t)  \\ \nonumber
&\qquad\qquad = f_{\lambda c}(t)\Theta(R-r)Y(\theta) ~ {V_k \over 8\pi^3 }   \\ \label{5}
&\qquad\qquad = n_{\lambda c}(t)\Theta(R-r)Y(\theta)   \nonumber
\end{flalign}
with
\begin{equation}
   f_{\lambda c}(t) = \frac{8\pi^2 }{ m_a^3\beta }  n_{\lambda c}(t)  \label{photon-rel} 
\end{equation}
where $f_\lambda(k, r, \theta, t)$ and $n_\lambda( r, \theta, t)$ are the photon occupation number and photon number density,  of helicity $\lambda=\pm1$ respectively, which are related by eq.(\ref{photon-rel}) and $V_k$ is the volume of the momentum spherical shell
\begin{flalign}\nonumber
V_k &= \int_{k_-}^{k_+} d^3k = 4\pi ({m_a\gamma \over2})^2 (k_+-k_-)= \pi m_a^3\beta\gamma^3 \approx \pi m_a^3\beta .
\end{flalign}

We assume that the number density of each helicity state is the same, so the total photon number density $n_\gamma$ can be written as
\begin{flalign}\nonumber
 n_\gamma( r, \theta, t) &= n_{\gamma c}(t)\Theta(R-r)Y(\theta)  = n_+( r, \theta, t) + n_-( r, \theta, t) &&\\ \nonumber
&= [n_{+c}(t)+n_{-c}(t)]\Theta(R-r)Y(\theta)  &&\\  
 n_{+c}(t) &= n_{-c}(t) \qquad n_{\gamma c}(t)=2n_{\lambda c}(t)  \label{lambda-gamma}
\end{flalign}
which defines $n_{xc}$.
Hence the coefficient of the total photon number density is just 2 times that of photon number density of each helicity state.

  The evolution relation between axion and photon occupation numbers is (see equation (13) of \cite{Kephart:1994uy})
\begin{flalign} \label{13Kephart}
&\frac{d f_\lambda ( k )  }{dt} =\frac{m_a \Gamma_a}{k^2} \int_{\frac{m_a^2}{4k}} d k_1     \\ \nonumber
&\times \{  f_a  ( k+k_1 ) [ 1+ f_\lambda ( k ) + f_\lambda ( k_1 ) ] - f_\lambda ( k )f_\lambda ( k_1 ) \}   
\end{flalign}
where $f_\lambda ( k )$ and $f_\lambda ( k_1 )$ are photon  occupation numbers of momentum $k$ and $k_1$, respectively. Other variables in $f_\lambda ( k )$ and $f_\lambda ( k_1 )$, i.e. $r, \theta, t$, are the same since they share the same spacetime. $ f_a  ( k+k_1 ) $ is the axion occupation number of momentum $k+k_1$. $\Gamma_a$ is the spontaneous axion decay rate.

This evolution equation can be integrated over $k$ and $k1$ phase space to yield (see the Appendix)

\begin{flalign} \nonumber
\frac{d n_\lambda  }{dt}  
&= \Theta(R-r)  \frac{m_a^3 \Gamma_a}{ 8\pi^2 } \bigg\{    [f_{ac} ( x + 2f_{\lambda c} \, xy    ) -f_{\lambda c}^2  y^2 ]   \times &&\\ \label{9}
&\quad [2\gamma^2\beta -\ln\bigg( \frac{1+\beta}{1-\beta} \bigg) ] - f_{\lambda c}^2  y^2 \times (2 \gamma^2\beta^2    ) \bigg\}  
\end{flalign}
Now we employ the nonrelativistic approximation ($\beta\ll1$).
\begin{flalign} \nonumber
& 2\gamma^2\beta = \frac{2\beta}{1-\beta^2} \approx 2\beta(1+\beta^2) = 2\beta + 2\beta^3    \\ \nonumber
& \ln\bigg( \frac{1+\beta}{1-\beta} \bigg) \approx 2\beta + \frac{2\beta^3}{3}      \\ \nonumber
& 2 \gamma^2\beta^2 \approx  2\beta^2 
\end{flalign}
to arrive at
\begin{flalign} \nonumber
&\frac{d n_\lambda  }{dt} =   \Theta(R-r)   \\ \nonumber
&  \times \frac{m_a^3 \Gamma_a  \beta^2  }{ 6\pi^2 } \bigg\{    f_{ac} ( X + 2f_{\lambda c} \, XY    )\beta - ( \beta + { 3\over2} )     f_{\lambda c}^2  Y^2     \bigg\}  
\end{flalign}

Substituting  the derived relations \eqref{axion-fn} and  \eqref{5}  into  \eqref{9}
\begin{flalign} \nonumber
\frac{d n_\lambda  }{dt} &= \Gamma_a \Theta(R-r)  &&\\ \nonumber 
&\qquad  \times\bigg[  n_{ac} ( X + \frac{ 16 \pi^2 n_{\lambda c} }{ \beta m_a^3 }  XY   )
- \frac{ 32 \pi^2 n_{\lambda c}^2 }{ 3 m_a^3 } (\beta + \frac{3}{2} ) Y^2  \bigg]  ~.  
\end{flalign}
Taking into consideration photon surface loss
\begin{flalign} \nonumber
\bigg( \frac{d n_\lambda  }{dt}  \bigg)_{\textrm{  surface loss}}  = - \frac{3c n_\lambda}{2R}
= - \frac{3c}{2R} n_{\lambda c} \Theta(R-r) Y(\theta)   ~,
\end{flalign}
we have an equation which gives the  number density for each helicity state
\begin{flalign} \nonumber
\frac{d n_\lambda  }{dt}  &=   \Theta(R-r) \times \bigg[  \frac{n_{ac}}{\tau_a} X(\theta) + 
\frac{ 16 \pi^2 n_{ac} n_{\lambda c} }{ \beta m_a^3 \tau_a }    X(\theta)  Y(\theta)     &&\\  \nonumber 
&\quad   - \frac{ 32 \pi^2 n_{\lambda c}^2 }{ 3 m_a^3 \tau_a } (\beta + \frac{3}{2} ) Y(\theta)^2  
- \frac{3c  n_{\lambda c} }{2R}  Y(\theta)   \bigg]     &&\\ \nonumber 
\end{flalign}
where we are assuming, as was shown in \eqref{lambda-gamma}, that total number density of photon is twice that of the individual helicity states. Therefore the rate of change of total number density of photon is
\begin{flalign} \nonumber
\frac{d n_\gamma  }{dt}  &=   \Theta(R-r) \times \bigg[  2 \frac{n_{ac}}{\tau_a} X(\theta) + 
\frac{ 16 \pi^2 n_{ac} n_{\gamma c} }{ \beta m_a^3 \tau_a }   X(\theta) Y(\theta)    &&\\  \nonumber 
&\quad - \frac{ 16 \pi^2 n_{\gamma c}^2 }{ 3 m_a^3 \tau_a } (\beta + \frac{3}{2} ) Y^2(\theta)  
- \frac{3c  n_{\gamma c} }{2R}  Y(\theta)   \bigg]     ~.
\end{flalign}

  Since from  \eqref{lambda-gamma}
\begin{flalign}\nonumber 
\frac{d n_\gamma  }{dt} & = \frac{d n_{\gamma c}  }{dt} \Theta(R-r) Y(\theta) ~,
\end{flalign}
if we drop the step function $\Theta(R-r)$
we have an equation for the coefficient of total number density of photon
\begin{flalign}\label{photoncoeff1}
\frac{d n_{\gamma c}  }{dt}  &= 2 \frac{n_{ac}}{\tau_a} \frac{X(\theta)}{Y(\theta)} + 
\frac{ 16 \pi^2 n_{ac} n_{\gamma c} }{ \beta m_a^3 \tau_a }    X(\theta)    &&\\  \nonumber
&\quad  - \frac{ 16 \pi^2 n_{\gamma c}^2 }{ 3 m_a^3 \tau_a } (\beta + \frac{3}{2} ) Y(\theta)  
- \frac{3c  n_{\gamma c} }{2R}   ~. 
\end{flalign}
From the first to the last term on the right hand side(RHS) of the equation, the terms account for spontaneous decay of axions, photon stimulated decay of axions, back reaction of photons, and surface loss of photons, respectively. Following similar approach, we  obtain an equation regarding the coefficient of total number density of axions
\begin{flalign}\label{axioncoeff1}
\frac{d n_{a c}  }{dt}  &= - \frac{n_{ac}}{\tau_a} \frac{X(\theta)}{Y(\theta)} - 
\frac{ 8 \pi^2 n_{ac} n_{\gamma c} }{ \beta m_a^3 \tau_a }    X(\theta)   
+ \frac{ 8 \pi^2 n_{\gamma c}^2 \beta }{ 3 m_a^3 \tau_a } Y(\theta)    ~.   
\end{flalign}

The third term on the RHS of \eqref{axioncoeff1} is proportional to $\beta$, while the third term on the RHS of \eqref{photoncoeff1} has a factor of $(\beta+{3\over2} )$.  Keeping track of two parts of axions generated from the back reacting  photons, we find that the ${3\over2}$ in the third term on the RHS of \eqref{photoncoeff1} represents sterile axions and it should have been and was excluded in the derivation of \eqref{axioncoeff1}.\\
\indent The left hand sides (LHS) of  \eqref{photoncoeff1} and \eqref{axioncoeff1} have no $\theta$ dependence, but the RHS does. $X(\theta)=Y(\theta)$ won't make \eqref{photoncoeff1} and \eqref{axioncoeff1} valid simultaneously. So even if there is some outside constraint which can keep the axions in the $X(\theta)$ distribution fixed, the photons cannot have the same distribution, i.e., $Y(\theta)\neq X(\theta)$.\\
\indent There is no simple way to find a closed form for $Y(\theta)$ because the LHS of the equations \eqref{photoncoeff1} and \eqref{axioncoeff1} have no $\theta$ dependence, while the $\theta$ dependences on the RHS of these equations are different. This suggests the possibility that  $Y(\theta)$ may be found as a series expansion in $X(\theta)$. As a first test of this idea we replaced the general form $X(\theta)$ with $\sin\theta$ to study the distribution with more axions accumulated near the equatorial plane with few near the polar area, aiming at matching orders of $\sin\theta$ on each side of equations. But this fails as it turns out that $\sin^n \theta ~ (n\in\mathbb{Z})$ is not an orthogonal set of functions and thus the calculation leads to contradictions. Therefore, we must expand the occupation numbers and number density in terms of a full set of orthogonal  functions. We do this in the next section where we choose the set to be the real spherical harmonics.
\section{Real sperical harmonics expansion}
The  set-up here is similar to the previous discussion except that the axion and photon occupation numbers and number densities have coefficients labeled by order index $l$ and $m$. For the axions
\begin{flalign}\nonumber
f_a(p, r, \Omega, t) =& \sum_{lm}f_{a lm}(t)Y_{lm}(\Omega)  \Theta(p_{\mbox{\tiny max}}-p)\, \Theta(R-r)    \\ \nonumber
  n_a(r, \Omega, t)=& \sum_{lm} n_{a lm}(t) Y_{lm}(\Omega)   \, \Theta(R-r)  \\ \nonumber
  f_{a lm}(t) =& \frac{6\pi^2  }{ m_a^3 \beta^3}  n_{a lm}(t)   \end{flalign}
where we have set
\begin{flalign}\nonumber
f_{ac}(t)X(\theta)= \sum_{lm} f_{a lm}(t) Y_{lm}(\Omega)
 \end{flalign}
Note that  $n_a$ can not be any superposition of real spherical harmonics, it  has to be real and positive, so it should be   put into the form
\begin{flalign} \nonumber
 n_a =& \Theta(R-r)   ( \sum_{l'm'} n_{a l' }^{m'}   Y_{l'}^{m'} )^*  (\sum_{lm} n_{a l }^m   Y_l^m  )  ~,
\end{flalign}
where $Y_l^m$ are complex spherical harmonics. This also applies to photons.
\begin{flalign}\nonumber
 f_\lambda(k, r, \Omega, t) =& \sum_{lm}f_{\lambda lm}(t)Y_{lm}(\Omega)       \\ \nonumber
&\times  \Theta(R-r) \Theta(k_+-k) \Theta(k-k_-)   \\ \nonumber
 n_\lambda( r, \Omega, t)  =& \sum_{lm}n_{\lambda lm}(t)Y_{lm}(\Omega)  \Theta(R-r)      \\ \nonumber
 f_{\lambda lm}(t)  =& \frac{8\pi^2  }{ m_a^3\beta } n_{\lambda lm}(t)  
\end{flalign}
\begin{flalign}\nonumber
 n_\gamma( r, \Omega, t)  
=&  \sum_{lm}[n_{+ lm}(t)+ n_{- lm}(t)]Y_{lm}(\Omega)  \Theta(R-r)     \\ \nonumber
=&    \sum_{lm}n_{\gamma lm}(t)Y_{lm}(\Omega)  \Theta(R-r)      \\ \nonumber
n_{+ lm}(t)  =&   n_{- lm}(t)      \qquad
n_{\gamma lm}(t)  = 2n_{\lambda lm}(t)
\end{flalign}
where similar the the axion case we have set
\begin{flalign}\nonumber
f_{\lambda c}(t)Y(\theta)= \sum_{lm} f_{\lambda lm}(t) Y_{lm}(\Omega).
 \end{flalign}
Following the steps from the previous general discussion, we have an equation similar to \eqref{photoncoeff1} for each choice of $lm$
\begin{flalign} \label{photoncoeff2}
 \frac{ d n_{\gamma lm}(t) }{ dt }  = &     2 {  n_{a lm}  \over  \tau_a  } + \frac{ 16 \pi^2  }{ \beta m_a^3 \tau_a  } E_{l m }       \\ \nonumber
&   - \frac{ 16 \pi^2  }{ 3 m_a^3  \tau_a } (\beta+ {3\over2} )   F_{l m }   
       -  \frac{3c}{2R}  n_{\gamma lm}(t)  ~,
\end{flalign}
where $E_{l m }$ and $F_{l m }$ are defined through
\begin{flalign} \label{elm}
n_a(\Omega, t)  n_{\gamma}(\Omega, t)   =&    \sum_{l'm'l'' m''}  n_{a l'm'}  n_{\gamma l'' m''}  Y_{l'm'} Y_{l''m''}   \\\nonumber
=&   \sum_{lm}  E_{l m } Y_{l m } 
\end{flalign}
and
\begin{flalign} \label{flm}
[n_{\gamma}(\Omega, t)]^2  =& \sum_{l'm'l'' m''}  n_{\gamma l'm'}  n_{\gamma l'' m''}  Y_{l'm'} Y_{l''m''} \\\nonumber
=&   \sum_{lm}  F_{l m } Y_{l m }   ~.
\end{flalign}

We also have equations similar to equation \eqref{axioncoeff1} for each choice of $lm$  with regard to the changing number density of axions. The equation includes components representing spontaneous decay, stimulated decay and back reaction with sterile axions excluded
\begin{flalign} \label{axioncoeff2} 
\frac{ d n_{a lm}(t) }{ dt }     &=   -  {  n_{a lm}  \over  \tau_a  }  -  \frac{ 8 \pi^2  }{ \beta m_a^3 \tau_a  } E_{l m }      
                                                + \frac{ 8 \pi^2 \beta }{ 3 m_a^3  \tau_a }   F_{l m }    ~.
\end{flalign}
The sterile axions evolve according to
\begin{flalign} \label{saxioncoeff2} 
\frac{ d n_{s lm}(t) }{ dt }     &=       \frac{ 4 \pi^2  }{  m_a^3  \tau_a }   F_{l m }    ~.
\end{flalign}
The rate of change of photon number density component can be expressed in terms of the changing components of normal axion and sterile axion, and the components of surface loss
\begin{flalign} \label{photoncoeff3}
 \frac{ d n_{\gamma lm}(t) }{ dt }  = &     -2 [ \frac{ d n_{a lm}(t) }{ dt }   + \frac{ d n_{s lm}(t) }{ dt } ]
       -  \frac{3c}{2R}  n_{\gamma lm}(t)  ~.
\end{flalign}

We now proceed to explore some example choices of initial axion distributions.
\section{Examples}
\subsection{$Y_{ 00 }$ distribution}
As a first example we consider the spherical symmetric axion distribution where 
the only nonzero component of axion number density is $n_{a 00}$,
\begin{flalign} \nonumber
 n_a =& \Theta(R-r)   n_{a 00 }   Y_{ 00 }(\Omega)     ~,
\end{flalign}
then
\begin{flalign} \nonumber
 n_{a lm}=0  \quad(lm\neq00)  ~.
\end{flalign}
This simplifies equation \eqref{elm} to
\begin{flalign} \nonumber
n_a(\Omega, t)  n_{\gamma}(\Omega, t)       =  &      \sum_{lm}  n_{a 00} Y_{00}   n_{\gamma lm}  Y_{lm} ~.
\end{flalign}
In addition, there is now a relationship between $E_{lm}$ and $n_{\gamma lm}$,
\begin{flalign} \label{00-elm-photonlm}
E_{lm}=n_{a 00} Y_{00}   n_{\gamma lm}  ~.
\end{flalign}
Equation \eqref{axioncoeff2} is also simplified for $lm\neq00$ to
\begin{flalign} \nonumber
0    &=  0   -  \frac{ 8 \pi^2  }{ \beta m_a^3 \tau_a  } E_{l m }      
                                                + \frac{ 8 \pi^2 \beta }{ 3 m_a^3  \tau_a }   F_{l m }     ~,
\end{flalign}
which reduces to
\begin{flalign} \label{00-elm-flm} 
 F_{l m }  &=      \frac{ 3  }{ \beta^2    } E_{l m }     \quad(lm\neq00)  ~.
\end{flalign}
Substitute \eqref{00-elm-photonlm} and \eqref{00-elm-flm}  into equation \eqref{flm} gives, upon splitting of $00$ pieces, the two forms of  \eqref{flm}
\begin{flalign} 
& [n_{\gamma}(\Omega, t)]^2      \label{00-ngng-1}
= F_{00} Y_{00} + \frac{ 3  }{ \beta^2    } n_{a 00} Y_{00}  \sum_{lm\neq00}  n_{\gamma l m } Y_{l m } \\ \nonumber
 F_{l m }  &=      \frac{ 3  }{ \beta^2    } E_{l m }     \quad(lm\neq00)  ~.
\end{flalign}
and
\begin{flalign} 
& [n_{\gamma}(\Omega, t)]^2  
= { n_{\gamma 00 } n_{\gamma 00 } \over 2\sqrt\pi}    Y_{00} +  2 n_{\gamma 00 }   Y_{00} \sum_{lm\neq00}  n_{\gamma l m } Y_{l m }     \nonumber \\
& + \sum_{l'm'l'' m'' \neq 0000}  n_{\gamma l'm'}  n_{\gamma l'' m''}  Y_{l'm'} Y_{l''m''}    ~.  \label{00-ngng-2}
\end{flalign}
The most conspicuous solution to the equation is
\begin{flalign} \nonumber 
F_{00} ={ n_{\gamma 00 } n_{\gamma 00 } \over 2\sqrt\pi}     ~,\qquad
n_{\gamma lm}   =0 \quad(lm\neq00)  ~.
\end{flalign}
where the only nonzero component of photon number density is $n_{\gamma 00}$. So if there is spherical symmetry in the axion distribution, then spherical symmetry also exist in photon distribution.

Now we  argue that this is the only solution of finite spherical harmonics series. Suppose that the highest spherical harmonics in the photon number density $n_{\gamma}(\Omega, t)$ is $Y_{l_a m_a}$. According to \eqref{00-ngng-1} and taking the $Y_{00}$ as a number, the highest spherical harmonics in $ [n_{\gamma}(\Omega, t)]^2 $ is also $Y_{l_a m_a}$. However, according to \eqref{00-ngng-2}, the highest spherical harmonics in $ [n_{\gamma}(\Omega, t)]^2 $ is going to be $Y_{2l_a \, 2m_a}$. This contradiction can only be resolved when $n_{\gamma lm}   =0 \ (lm\neq00)$, i.e. the photon number density retains spherical symmetry.\\

The reason why this is the only finite series case is that the $Y_{00}$ distribution of axions mathematically requires the photons to couple in a specific way that retains the $Y_{00}$ distribution of axions, as is implied by equations \eqref{00-elm-photonlm} and \eqref{00-elm-flm}. 

Now we know  all the coupling coefficients $E_{l m}$ and $F_{l m}$,
\begin{flalign} \nonumber 
E_{l m}={ n_{a 00 } n_{\gamma 00 } \over 2\sqrt\pi} \delta_{l0}\delta_{m0}     ~,\qquad  F_{lm} ={ n_{\gamma 00 } n_{\gamma 00 } \over 2\sqrt\pi}  \delta_{l0}\delta_{m0}   ~.
\end{flalign}
Equations \eqref{photoncoeff2}, \eqref{axioncoeff2} and \eqref{saxioncoeff2}  reduce to the equations (34'), (37'), (38') in \cite{Kephart:1994uy} given that
\begin{flalign} \nonumber 
n_{\gamma 00 } =  2\sqrt\pi  n_{\gamma  }     ~,\qquad  n_{a 00 } =  2\sqrt\pi  n_{a  }      ~,
\end{flalign}
because it is the $n_{\gamma 00 }Y_{00}$ that describes the photon number density. Hence we have checked the spherically symmetric model results given in  \cite{Kephart:1994uy} .

\subsection{$Y_{ 20 }$ distribution}
For a $Y_{ 20 }$ axion distribution the only nonzero component of the axion number density is $n_{a 20}$,
\begin{flalign} \nonumber
 n_a =& \Theta(R-r)   n_{a 20 }   Y_{ 20 }(\Omega)     ~,
\end{flalign}
so that
\begin{flalign} \nonumber
 n_{a lm}=0  \quad(lm\neq20)  ~.
\end{flalign}
Equation \eqref{axioncoeff2} is  simplified for $lm\neq20$, to
\begin{flalign} \nonumber
0    &=  0   -  \frac{ 8 \pi^2  }{ \beta m_a^3 \tau_a  } E_{l m }      
                                                + \frac{ 8 \pi^2 \beta }{ 3 m_a^3  \tau_a }   F_{l m }     ~,
\end{flalign}
which reduces to
\begin{flalign} \label{20-elm-flm} 
 F_{l m }  &=      \frac{ 3  }{ \beta^2    } E_{l m }     \quad(lm\neq20)  ~.
\end{flalign}
The  nonzero component $ n_{a 20}$ of axion number density evolves via
\begin{flalign}  \nonumber
\frac{ d n_{a 20}(t) }{ dt }     &=   -  {  n_{a 20}  \over  \tau_a  }  -  \frac{ 8 \pi^2  }{ \beta m_a^3 \tau_a  } E_{20}      
                                                + \frac{ 8 \pi^2 \beta }{ 3 m_a^3  \tau_a }   F_{20}    ~.
\end{flalign}
The photon number density component $ n_{\gamma 20}$ growth rate is 
\begin{flalign} \nonumber
 \frac{ d n_{\gamma 20}(t) }{ dt }  = &     2 {  n_{a 20}  \over  \tau_a  } + \frac{ 16 \pi^2  }{ \beta m_a^3 \tau_a  } E_{20}       \\ \nonumber
&   - \frac{ 16 \pi^2  }{ 3 m_a^3  \tau_a } (\beta+ {3\over2} )   F_{20 }   
       -  \frac{3c}{2R}  n_{\gamma 20}(t)  ~,
\end{flalign}
while the other photon number density component $ n_{\gamma lm}(lm\neq20)$  evolve as
\begin{flalign} \nonumber
 \frac{ d n_{\gamma lm}(t) }{ dt }  =    - \frac{ 8 \pi^2  }{   m_a^3  \tau_a }     F_{lm }   
       -  \frac{3c}{2R}  n_{\gamma lm}(t)  ~.
\end{flalign}
Since no spontaneous decay from axion feeds into these components, they are negligible. This example is not physical because a number density of the form $Y_{ 20 }$ becomes negative in some regions. It is included here for demonstration purpose. The next examples is physical and motivated by superradience.

\subsection{$  Y_1^{\pm1*} Y_1^{\pm1}\sim\sin^2\theta$ distribution}
A $sin^2\theta$ distribution is torodial  and is positive definite everywhere, and hence can represent a physical distribution of particles. For this case
the only nonzero components of the axion number density are $n_{a 00}$ and $n_{a 20}$,
so we can write $ n_a(r,\theta,t)$ in several useful forms
\begin{flalign} \nonumber
 n_a =& \Theta(R-r)   n_a(t)  \sin^2\theta       \\ \nonumber
=& \Theta(R-r)  n_a(t)  \frac{4\sqrt\pi}{3} ( Y_{00}-\frac{1}{\sqrt5}Y_{20})      \\ \nonumber
=&\Theta(R-r)  [ n_{a 00}(t)   Y_{00}  + n_{a 20}(t)   Y_{20} ] ~.
\end{flalign}
The relation between $n_{a 00}$ and $n_{a 20}$ is
\begin{flalign} \label{0020-axioncoeff}
   n_{a 20}(t)    =  - {  n_{a 00}(t)  \over \sqrt5 } ~.
\end{flalign}
 Similar to previous examples, we find that for components other than $00$ and $20$
\begin{flalign}  \nonumber 
 F_{l m }  &=      \frac{ 3  }{ \beta^2    } E_{l m }     \quad(lm\neq00,20)  ~,
\end{flalign}
so that the components of photon number density evolve as
\begin{flalign} \nonumber
 \frac{ d n_{\gamma lm}(t) }{ dt }  =    - \frac{ 8 \pi^2  }{   m_a^3  \tau_a }     F_{lm }   
       -  \frac{3c}{2R}  n_{\gamma lm}(t)  ~.
\end{flalign}\\

Since no spontaneous decay from axion feeds into these components, they are negligible, as  in the previous example. The nonzero axion number density components are given by
\begin{flalign}  \nonumber
\frac{ d n_{a 00}(t) }{ dt }   +  {  n_{a 00}  \over  \tau_a  }  &=     -  \frac{ 8 \pi^2  }{ \beta m_a^3 \tau_a  } E_{00}      
                                                + \frac{ 8 \pi^2 \beta }{ 3 m_a^3  \tau_a }   F_{00}    \\  \nonumber
\frac{ d n_{a 20}(t) }{ dt }   + {  n_{a 20}  \over  \tau_a  }    &=   -  \frac{ 8 \pi^2  }{ \beta m_a^3 \tau_a  } E_{20}      
                                                + \frac{ 8 \pi^2 \beta }{ 3 m_a^3  \tau_a }   F_{20}    ~.
\end{flalign}
Because of \eqref{0020-axioncoeff}, this leads to the relation
\begin{flalign}  \nonumber
&-  \frac{ 8 \pi^2  }{ \beta m_a^3 \tau_a  } E_{20}  + \frac{ 8 \pi^2 \beta }{ 3 m_a^3  \tau_a }   F_{20}     \\  \label{0020-elm-flm}
=&       {1\over\sqrt5}    (  \frac{ 8 \pi^2  }{ \beta m_a^3 \tau_a  } E_{00}  -  \frac{ 8 \pi^2 \beta }{ 3 m_a^3  \tau_a }   F_{00}   )       .
\end{flalign}
The photon number density component $ n_{\gamma 20}$ grows as
\begin{flalign} \nonumber
 \frac{ d n_{\gamma 00}(t) }{ dt }  = &     2 {  n_{a 00}  \over  \tau_a  } + \frac{ 16 \pi^2  }{ \beta m_a^3 \tau_a  } E_{00}       \\ \nonumber
&   - \frac{ 16 \pi^2  }{ 3 m_a^3  \tau_a } (\beta+ {3\over2} )   F_{00 }   
       -  \frac{3c}{2R}  n_{\gamma 00}(t)   \\ \nonumber
       \end{flalign}
       and
       \begin{flalign} \nonumber
 \frac{ d n_{\gamma 20}(t) }{ dt }  = &     2 {  n_{a 20}  \over  \tau_a  } + \frac{ 16 \pi^2  }{ \beta m_a^3 \tau_a  } E_{20}       \\ \nonumber
&   - \frac{ 16 \pi^2  }{ 3 m_a^3  \tau_a } (\beta+ {3\over2} )   F_{20 }   
       -  \frac{3c}{2R}  n_{\gamma 20}(t)  ~.
\end{flalign}
Because of \eqref{0020-axioncoeff} and \eqref{0020-elm-flm}, we can combine the previous two equations and write
\begin{flalign} \nonumber
& \frac{ d n_{\gamma 00}(t) }{ dt }  + \frac{3c}{2R}  n_{\gamma 00}(t)   \\ \nonumber
  = &     2 {  n_{a 00}  \over  \tau_a  } + \frac{ 16 \pi^2  }{ \beta m_a^3 \tau_a  } E_{00}         - \frac{ 16 \pi^2  }{ 3 m_a^3  \tau_a } (\beta+ {3\over2} )   F_{00 }       ~,     \\ \nonumber
& \frac{ d n_{\gamma 20}(t) }{ dt }  + \frac{3c}{2R}  n_{\gamma 20}(t)   \\ \nonumber
= &     2 {  n_{a 00}  \over  \tau_a  } ({-1\over\sqrt5}) +  (\frac{ 16 \pi^2  }{ \beta m_a^3 \tau_a  } E_{00}         - \frac{ 16 \pi^2  \beta }{ 3 m_a^3  \tau_a }   F_{00 }   )   ({-1\over\sqrt5})  \\ \nonumber
  & - \frac{ 8 \pi^2  }{  m_a^3  \tau_a }     F_{20 }   \\ \nonumber
= &    {-1\over\sqrt5}  [ \frac{ d n_{\gamma 00}(t) }{ dt }  + \frac{3c}{2R}  n_{\gamma 00}(t) ]  - \frac{ 8 \pi^2  }{  m_a^3  \tau_a }     F_{20 }  
\end{flalign}
We observe that if the part of back reaction that results in sterile axions is neglected, then
\begin{flalign} \nonumber n_{\gamma 20}(t)    =  - {  n_{\gamma 00}(t)  \over \sqrt5 }   ~, \end{flalign}
so the photons would remain in $\sin^2\theta$ distribution.\\

\subsection{General distribution}
Suppose that we have an axion number density
\begin{flalign} \nonumber
 n_a =& \Theta(R-r)  \sum    n_{a l m }   Y_{ l m }(\Omega)     ~,
\end{flalign}

For $n_{a l m } = 0$, then according to \eqref{axioncoeff2} this leads to
\begin{flalign} \nonumber
F_{lm} =& { 3 \over \beta^2 }  E_{lm}  ~~  ( n_{a l m } = 0 ) 
\end{flalign}
Substituting this condition into equation \eqref{photoncoeff2}, we have
\begin{flalign} \nonumber
 \frac{ d n_{\gamma lm}(t) }{ dt }  = &         - \frac{ 16 \pi^2  }{ 3 m_a^3  \tau_a } ( {3\over2} )   F_{l m }   
       -  \frac{3c}{2R}  n_{\gamma lm}(t)    
\end{flalign}
also for  $n_{a l m } = 0$. 
Hence there is no source feeding those photon components.

The parts of back reaction that results in sterile axions and surface loss are the only terms that contribute to these components. It is expected that these components die out quickly and thus have no effect on lasing. So
\begin{flalign} \nonumber
n_{\gamma} =  \,    & \Theta(R-r)  \sum   n_{\gamma l m  }   Y_{ l  m }(\Omega)     ~.
\end{flalign}
where
\begin{flalign} \nonumber
n_{\gamma l m } \approx   0   ~~  ( \text{when } n_{a l m } = 0 ) ~.
\end{flalign}
I.e., the photon field has the same spherical harmonic components as the axion field, as other components  die out quickly due to lack of sources. Neither spontaneous decay nor stimulated decay contributes to the harmonic components of photons that are not present in the axions.
 
Suppose that all the axion components are nonzero, and they are proportional to each other,
$$n_{a l  m } = \alpha_{lm} n_{a l_0 m_0 } ,$$  where $\alpha_{lm}$ are numbers and $n_{a l_0 m_0 }$ is the fiducial component to which all other components are proportional. Then
\begin{flalign} \nonumber
\frac{ 8 \pi^2 \beta }{ 3 m_a^3  \tau_a }   F_{ lm }   -  \frac{ 8 \pi^2  }{ \beta m_a^3 \tau_a  } E_{ lm }  =    \frac{ d n_{a lm }(t) }{ dt }     +    {  n_{a lm  }  \over  \tau_a  }          \\ \nonumber
\end{flalign}
and
\begin{flalign} \nonumber
& \frac{ d n_{\gamma  lm }(t) }{ dt } +   \frac{3c}{2R}  n_{\gamma  lm  }(t)           \\ \nonumber
= &     2 {  n_{a   lm  }  \over  \tau_a  } + \frac{ 16 \pi^2  }{ \beta m_a^3 \tau_a  } E_{  lm  }      
          - \frac{ 16 \pi^2  }{ 3 m_a^3  \tau_a } (\beta+ {3\over2} )   F_{   lm   }      \\ \nonumber
=&    -2  \frac{ d n_{a  lm }(t) }{ dt }  - \frac{ 8 \pi^2  }{  m_a^3  \tau_a }     F_{   lm }  
\end{flalign}
If the part of the back reaction that results in sterile axions is neglected, then
\begin{flalign} \nonumber
& \frac{ d n_{\gamma  lm  }(t) }{ dt } +   \frac{3c}{2R}  n_{\gamma  lm  }(t)           \\ \nonumber
=&    -2  \frac{ d n_{a  lm  }(t) }{ dt }   =   -2  \alpha_{ lm } \frac{ d n_{a  l_0 m_0 }(t) }{ dt }    \\ \nonumber
=& \alpha_{ lm }  [   \frac{ d n_{\gamma  l_0 m_0 }(t) }{ dt } +   \frac{3c}{2R}  n_{\gamma  l_0 m_0 }(t)   ]
\end{flalign}
\begin{flalign} \nonumber
n_{\gamma   lm }= \alpha_{ lm }  n_{\gamma  l_0 m_0 }   \end{flalign}
Hence the distribution of photons would keep the same shape as that of the axions if sterile axions were neglected.\\

\section{Discussion}
The calculation presented here tells one the initial spatial distribution of photons once the spatial distribution of the axions is given. It does not give direct instructions on how to achieve observable effects from axion cluster lasing. The model does take the mechanism that the stimulated decay of axion produces type of photons that have the same momenta as the photons which induced the stimulated decay process.
\begin{flalign} \nonumber
 2k \frac{d f_\lambda ( \vec{k} )  }{dt}  =&  \frac{4 m_{a} \Gamma_{a} }{ \pi }  \int \frac{d^3  k_1   }{ 2k_1^0 } \frac{d^3  p   }{ 2p^0 }  \delta^4 (p-k-k_1) \times \\  \nonumber
   &  \{  f_a  ( \vec{p} ) [ 1+ f_\lambda ( \vec{k} ) + f_\lambda ( \vec{k}_1 ) ] - f_\lambda ( \vec{k} )f_\lambda ( \vec{k}_1 ) \}  ~.
\end{flalign}
However, there is a compromise made here by using this equation. The entire model is a local theory. The photon occupation number here and now depends only on particle occupation numbers here and now. If the cluster in the model is a ball and all the quantities are spherical symmetric, the local theory provides useful predictions about the lasing process. However, if the cluster is of some specific geometrical shape, then the local theory probably won't give pertinent information that reflect the geometry of the cluster. Thus we suggest that a non-local lasing theory which could be governed by the following equation,
\begin{flalign} \nonumber
 & 2k \frac{d f_\lambda ( \vec{k} ,\vec{x}, t )  }{dt} =   \frac{4 m_{a} \Gamma_{a} }{ \pi }  \int \frac{d^3  k_1   }{ 2k_1^0 } \frac{d^3  p   }{ 2p^0 }  \delta^4 (p-k-k_1)   \\  \nonumber
& \times  C \int d^3x' \{  f_a  ( \vec{p} ,\vec{x}', t' )   \int^t dt' e^{- \Gamma_{a}(t-t')} \delta[\vec{x} - \vec{x}' - {c\vec{k} \over k} (t-t')]    \\  \nonumber
   &+ f_a  ( \vec{p} ,\vec{x}, t )  \int^t dt' e^{- \Gamma_{a}(t-t')} \delta[\vec{x} - \vec{x}' - {c\vec{k} \over k} (t-t')] 
  f_\lambda ( \vec{k} ,\vec{x}' ,t' )    \\  \nonumber
& +f_a  ( \vec{p} ,\vec{x}, t )   \int^t dt' e^{- \Gamma_{a}(t-t')} \delta[\vec{x} - \vec{x}' - {c\vec{k}_1 \over k_1} (t-t') ]  
  f_\lambda ( \vec{k}_1 ,\vec{x}' ,t' )    \\  \nonumber
&- f_\lambda ( \vec{k} ,\vec{x}, t   )   \int^t dt' e^{- \Gamma_{a}(t-t')} \delta[\vec{x} - \vec{x}' - {c\vec{k}_1 \over k_1} (t-t')]
f_\lambda ( \vec{k}_1 ,\vec{x}' ,t' ) \} ~.
\end{flalign}
In the non-local model, the photon occupation number here and now depends on all the past occupation number of events that are casually connected to here and now.
The factor $e^{- \Gamma_{a}(t-t')}$ takes account the probability that photons propagating from $\vec{x}' $ to $\vec{x}$ without stimulating axion or going to annihilation.
\section{Appendix}
Starting from the  evolution relation between axion and photon occupation numbers \cite{Kephart:1994uy}
\begin{flalign} \label{13Kephart}
&\frac{d f_\lambda ( k )  }{dt} =\frac{m_a \Gamma_a}{k^2} \int_{\frac{m_a^2}{4k}} d k_1     \\ \nonumber
&\times \{  f_a  ( k+k_1 ) [ 1+ f_\lambda ( k ) + f_\lambda ( k_1 ) ] - f_\lambda ( k )f_\lambda ( k_1 ) \}   
\end{flalign}
where $f_\lambda ( k )$ and $f_\lambda ( k_1 )$ are photon  occupation numbers of momentum $k$ and $k_1$, respectively. Other variables in $f_\lambda ( k )$ and $f_\lambda ( k_1 )$, i.e. $r, \theta, t$, are the same since they share the same spacetime. $ f_a  ( k+k_1 ) $ is the axion occupation number of momentum $k+k_1$. $\Gamma_a$ is the spontaneous axion decay rate. Substitute \eqref{axionON} and \eqref{photonON} into \eqref{13Kephart} we arrive at
\begin{flalign}\nonumber
&\frac{d f_\lambda ( k )  }{dt} =\frac{m_a \Gamma_a}{k^2}  \bigg\{ [ 1+ f_\lambda ( k ) ]  f_{ac} \, \Theta(R-r)  X(\theta)  &&\\ \nonumber
&\quad \times \int_{\frac{m_a^2}{4k}} \Theta(p_{\textrm{\tiny max}}-\sqrt{(k+k_1)^2-m_a^2}) d k_1  &&\\ \nonumber
&\quad +f_{ac} f_{\lambda c} \, [\Theta(R-r)]^2  X(\theta) \, Y(\theta)  \,\times&&\\ \nonumber
&\int_{\frac{m_a^2}{4k}} \Theta( p_{\textrm{\tiny max}}-\sqrt{(k+k_1)^2-m_a^2})\Theta(k_+-k_1) \Theta(k_1-k_-) d k_1  &&\\ \nonumber
&-f_\lambda(k)  f_{\lambda c} \Theta(R-r) \, Y(\theta)  \int_{\frac{m_a^2}{4k}} \Theta(k_+-k_1) \Theta(k_1-k_-) d k_1  
\bigg\}  ~.
\end{flalign}
The first and second  integrals are the same,
\begin{flalign} \nonumber
&\quad  \int_{\frac{m_a^2}{4k}} \Theta(p_{\textrm{\tiny max}}-\sqrt{(k+k_1)^2-m_a^2}) d k_1  \\ \nonumber
&=\int_{\frac{m_a^2}{4k}} \Theta( p_{\textrm{\tiny max}}-\sqrt{(k+k_1)^2-m_a^2})\Theta(k_+-k_1) \Theta(k_1-k_-) d k_1  
\\ \nonumber
&=m_a\gamma-k-\frac{m_a^2}{4k} ~.
\end{flalign}
The third integral is related to the back reaction of photons. It is convenient to split it into two parts
\begin{flalign} \nonumber
&\quad  \int_{\frac{m_a^2}{4k}} \Theta(k_+-k_1) \Theta(k_1-k_-) d k_1     \\ \nonumber 
&=\int_{\frac{m_a^2}{4k}}^{m_a\gamma-k} d k_1 + \int_{m_a\gamma-k}^{k_+}   d k_1  \\ \nonumber 
&= (m_a\gamma-k-\frac{m_a^2}{4k} ) +  (k-k_-)    ~.
\end{flalign}
The first part represents back reaction resulting in axions with energy a  less than $m_a\gamma$ that  axions that can again participate in stimulated emission, while the second part gives the back reaction resulting in sterile axions, i.e., where the total energy of the axion $k+k_1$ is larger than $m_a\gamma$.

Moving the step function $\Theta(R-r)$ in front of the  curly brackets and substituting the results of the integrations, we have
\begin{flalign} \nonumber
 \frac{d f_\lambda ( k )  }{dt} &=  \Theta(R-r) \, \frac{m_a \Gamma_a}{k^2}  \bigg\{   &&\\ \nonumber 
&\quad [ 1+ f_\lambda ( k ) ]  f_{ac} \, X(\theta) \, (m_a\gamma-k-\frac{m_a^2}{4k} )  &&\\ \nonumber 
&\quad + f_{ac} f_{\lambda c} \,  X(\theta)  \, Y(\theta) \, (m_a\gamma-k-\frac{m_a^2}{4k} ) &&\\ \nonumber
&\quad -f_\lambda(k) f_{\lambda c}  Y(\theta) \, [ (m_a\gamma-k-\frac{m_a^2}{4k} ) +  (k-k_-) ]   \bigg\}.
\end{flalign}
Collecting  terms $f_\lambda(k)$ can be written
\begin{flalign} \nonumber
\frac{d f_\lambda ( k )  }{dt} &= \Theta(R-r) \Theta(k_+-k) \Theta(k-k_-) \, \frac{m_a \Gamma_a}{k^2}  \bigg\{   &&\\ \nonumber
&\qquad [f_{ac} ( X + 2f_{\lambda c} \, XY    ) -f_{\lambda c}^2  Y^2 ] (m_a\gamma-k-\frac{m_a^2}{4k} )  &&\\  \nonumber 
&\qquad -f_{\lambda c}^2  y^2  (k-k_-)
\bigg\}  ~.
\end{flalign}
The rate of change of photon number density is the integration of this equation over $k$ space
\begin{flalign} \nonumber
\frac{d n_\lambda  }{dt}  &= \int \frac{d f_\lambda ( k )  }{dt} \frac{d^3k}{(2\pi)^3}   &&\\ \nonumber
&= \Theta(R-r)  \frac{m_a \Gamma_a}{ 2\pi^2 } \bigg\{    [f_{ac} ( X + 2f_{\lambda c} \, XY    ) -f_{\lambda c}^2  Y^2 ]   \times &&\\ \nonumber
&\quad \int_{k_-}^{k_+} (m_a\gamma-k-\frac{m_a^2}{4k} ) dk - f_{\lambda c}^2  y^2  \int_{k_-}^{k_+}  (k-k_-) dk \bigg\}  
\end{flalign}
Evaluating the two integrals,
\begin{flalign} \nonumber
 & \int_{k_-}^{k_+} (m_a\gamma-k-\frac{m_a^2}{4k} ) dk
= \frac{m_a^2\gamma^2\beta}{2}  -\frac{m_a^2}{4}\ln\bigg( \frac{1+\beta}{1-\beta} \bigg)   \\ \nonumber
& \int_{k_-}^{k_+}  (k-k_-) dk = \frac{m_a^2\gamma^2\beta^2}{2}   ~.
\end{flalign}
gives
\begin{flalign} \nonumber
\frac{d n_\lambda  }{dt}  
&= \Theta(R-r)  \frac{m_a^3 \Gamma_a}{ 8\pi^2 } \bigg\{    [f_{ac} ( x + 2f_{\lambda c} \, xy    ) -f_{\lambda c}^2  y^2 ]   \times &&\\ \nonumber
&\quad [2\gamma^2\beta -\ln\bigg( \frac{1+\beta}{1-\beta} \bigg) ] - f_{\lambda c}^2  y^2 \times (2 \gamma^2\beta^2    ) \bigg\}  
\end{flalign}

\vfill

\vspace{0.2cm} 
\begin{acknowledgments}

\end{acknowledgments}

\end{document}